\documentclass[twocolumn,showpacs,preprintnumbers,amsmath,amssymb]{revtex4}
\usepackage[dvips]{graphicx}
\usepackage{amssymb,amsfonts,amsmath}

\begin{document}

\title{The microscopic meaning of grand potential resulting from combinatorial approach to a general system of particles}

\author{Agata Fronczak}
\email{agatka@if.pw.edu.pl}
\affiliation{Faculty of Physics, Warsaw University of Technology,Koszykowa 75, PL-00-662 Warsaw, Poland}

\date{\today}

\begin{abstract}
We present a completely new approach to the problem of interacting fluids, which we believe may provide important insights into microscopic mechanisms that lead to the occurrence of phase transitions. The approach exploits enumerative properties and combinatorial meaning of Bell polynomials. We derive the exact formula for probability of a general system of $N$ particles at temperature $T$ to consist of $k$ weakly coupled clusters of various sizes. We also show that the grand potential of the system may be considered as the exponential generating function for the number of internal states (thermodynamic probability) of these clusters. The microscopic interpretation of the grand potential is novel and surprising, especially if one recalls that until now the only thermodynamic meaning of this free energy was known. We also derive an approximated expression for the density of states. \end{abstract}


\maketitle

\section{Introduction and motivation}\label{sec0}

This work springs from a fascination with the concepts of equilibrium, interaction, and collective behavior in physical systems. The analyzed systems are considered to be made up of separate clusters which are composed of particles (elements, units etc.) interacting via short-range potentials. Exemplary systems where the assumptions apply are imperfect gases, for which the total interparticle potential energy is given by the sum of pair interactions characterized by a hard core and a weak attractive region. In this study we assume a particular type of interaction between clusters termed weak coupling. Weak coupling means that the only interaction between clusters is that provided by the migration of particles between them. For this reason, within the framework of statistical mechanics, the system can be actually described as consisting of disjoint, non-interacting, and independent clusters. A systematic method of dealing with such systems was first developed by Ursell \cite{1927Ursell}, whose work was later extended by Mayer \cite{1937MayerJChemPhys1, 1937MayerJChemPhys2, 1938MayerJChemPhys, 1941MayerJChemPhys, bookMayer, bookHill}, due to whom are the major advances in this field.

With this paper we start a new line of theoretical research on imperfect gases (or interacting fluids). Our approach originates in basic concepts of statistical physics (ensemble theory) \cite{bookReichl} and enumerative combinatorics \cite{bookStanley}. In particular, we derive a new, exact and general combinatorial formula for probability of a system of $N$ particles at temperature $T$ to consist of $k$ clusters of various sizes. To achieve the result we use the generating function for Bell polynomials \cite{bookComtet}. To interpret it we exploit combinatorial meaning of this function as given by the so-called exponential formula \cite{bookWilf}, which is the cornerstone of the art of counting in enumerative combinatorics.

To be definite, the exponential formula deals with the question of counting composite structures that are built out of a given set of building blocks \cite{bookWilf} (in the exemplary case of imperfect gases, the term "composite structure" stands for "microstate" of the system, while the term "block" means "cluster"). If you know how many blocks of each size there are, the formula tells you how many composite structures of each size can be built from those blocks. In short, the exponential formula states that the exponential generating function, $F(x)=\sum_{n=0}^\infty f_nx^n/n!$, for composite structures is the exponential of that for building blocks, $f(x)$, i.e. $F(x)=\exp[f(x)]$. With this statement in mind, the grand partition function in statistical physics \footnote{All the symbols used here have their usual meanings described later in the text.}, $\Xi(\beta,z)=\exp[-\beta\Phi(\beta,z)]$, which is also a kind of generating function for microscopic arrangements of the system investigated, gains a completely new interpretation as the generating function for composite states having a well-defined internal (clustered) structure. Furthermore, properties of this internal structure are encoded in another generating function, $\Phi(\beta,z)$, which appears to be free energy of the system.

The outlined understanding of the grand partition function as given by the exponential formula is novel, but its theoretical basis is partially covered by the cluster expansion method, set out over 50 years ago in the already mentioned works of Mayer \cite{1941MayerJChemPhys, bookMayer}. Mayer's ideas were later developed by others \cite{1938Born,1938Kahn}, for example by the distinguished physicist Uhlenbeck and his student Riddel, in whose thesis the exponential formula first appeared \cite{1953Riddel}. However, following initial interest among physicists, the idea of enumerative uses of generating functions was generalized and extended mainly by mathematicians (cf.~\cite{bookStanley,bookWilf}), and the physical origins of the concept have almost been forgotten.

This happens because the Mayer's original approach to imperfect gases failed to match real systems. Foremost it failed in the description of condensation effects in dense interacting gases for which it was originally developed. Thus, is it worth to coming back to this unsuccessful story? We believe, it is wort it, especially that given the present comprehension of different enumerative methods, one can easy trace the Mayer's failure and hopefully explain the controversy over rather subtle mathematical points involved in his condensation and critical point theory. It would be the great success of equilibrium statistical mechanics, because although the alternative formalism of Yang and Lee \cite{1952YangPhysRev} makes possible a simple mathematical description of condensation phenomena but it has still (after more than six decades of research) the \emph{disadvantage of being less detailed and physically rather obscure} (see \cite{bookHill}, p.~123) in comparison with the Mayer's approach.

Despite of its strong connections with the Mayer's theory, the ideas described in this paper were  developed entirely independently. For the reason, the connections are not easily seen at first glance. In fact, for a long time of struggling with the exponential formula, Bell polynomials etc., we were unaware of the previous contributions in the field. Now, after several months of rummaging through literature on the topic, we can locate our studies as much more general than the Mayer's approach, and as being complementary to Markov models of systems in stochastic equilibrium studied by Whittle \cite{bookWhittle}, and nowadays independently developed within the framework of stochastic combinatorial processes \cite{bookPitman, 2007BerestyckiJStatPhys}.

The work reported in this paper is quite general and fairly abstract. Our main result is that the grand potential may be considered as the exponential generating function for the number of internal states of clusters of different sizes. It puts the grand potential (also called the Landau free energy), $\Phi(\beta,z)$, in a completely new perspective. Up to now the only macroscopic interpretation of this thermodynamic state function was known and considered (see Sec.~2F5 in \cite{bookReichl}). The ideas advanced in this paper (and the forthcoming ones) reveal its microscopic meaning. This, in turn, may shed new light on microscopic origins of phase transitions, since they are traditionally classified according to the lowest-order derivative of the thermodynamic potential which changes discontinuously (or diverges) at the transition point (see Chaps. 3 and 7 in \cite{bookReichl}). In subsequent papers we will apply the methods here outlined to lattice gasses. We will also show how the virial expansion \cite{bookHill, bookPathria}of the equation of state for imperfect gases can be simply derived from our approach.

This paper is organized as follows. In the next section, Sec.~\ref{sec1}, the main result of this paper, i.e. the exact combinatorial formula is derived for thermodynamic probability of a general system of $N$ interacting particles to consists of $k$ weakly coupled clusters. Then, in the subsequent section, Sec.~\ref{sec2}, properties of the formula are discussed, the general integral expression for the density of states is introduced, the hypothesis about microscopic meaning of the grand potential is advanced, and the example of the ideal gas is worked out. Some general conclusions are given in the end of the paper.

\section{Derivation of the main result}\label{sec1}

In this paper, we limit ourselves to the simplest possible case: a system composed of identical monoatomic molecules without internal degrees of freedom. The thermodynamical state of the system is given by the temperature, $T$, and the chemical potential per molecule, $\mu$. We assume that the classical treatment is adequate and use the formalism of the grand canonical ensemble (see Sec.~7G in \cite{bookReichl}) to describe the open system in the infinite volume limit.

To derive the main result, we write the grand partition function (GPF) of a general system of particles in two ways. Firstly, we use a slightly modified textbook approach which exploits the definition of DOS as the number of states whose energies lie in some narrow range, between $E$ and $E+dE$. Secondly, we make use of the generating function for partial Bell polynomials \cite{bookComtet}, which has a very nice combinatorial interpretation given by the exponential formula \cite{bookWilf}. Then, we compare both expressions. For the reasons of clarity GPF is denoted, respectively, by $\Xi_1$ and $\Xi_2$, in the two approaches.

In the infinite volume limit the grand partition function, $\Xi_1(\beta,z)$,
of a general system of particles is defined as the sum of Boltzmann
exponential factors, $e^{-\beta(E-\mu N)}$, over all microscopic realizations
(microstates), $\Omega$, of the system
\begin{equation}\label{Xi1}
\Xi_1(\beta,z)=\sum_{\Omega}e^{-\beta(E(\Omega)-\mu N(\Omega))}=1+\!\!\sum_{N=1}^\infty z^N Z(\beta,N),
\end{equation}
where $\beta=(k_BT)^{-1},\mu,E$ and $N$, have their usual meanings
\footnote{i.e. they represent, respectively, the Boltzmann's constant, $k_B$,
temperature, $T$, chemical potential, $\mu$, energy, $E$, and the number of
particles, $N$.}, the parameter $z=e^{\beta\mu}$ is called fugacity (or
activity), and
\begin{equation}\label{Z1}
Z(\beta,N)=\sum_{\Omega}e^{-\beta E(\Omega)}=\int_0^\infty g(E,N)e^{-\beta E} dE
\end{equation}
stands for the canonical partition function (PF). Inserting Eq.~(\ref{Z1})
into Eq.~(\ref{Xi1}) and then multiplying the resulting expression by the
sum over the probability distribution, $f(k,E)$, that the considered system having energy $E$ consists of $k$ weakly coupled (i.e. statistically independent disjoint groups of particles)
\begin{equation}\label{af1}
\sum_{k=1}^N f(k,E)=h(E)\equiv 1,
\end{equation}
one gets
\begin{align}\nonumber
&\Xi_1(\beta,z)=\\&=1+\!\!\sum_{N=1}^\infty z^N \!\!\int_0^\infty\!\! g(E,N)e^{-\beta E}\left(\sum_{k=1}^N f(k,E)\right)dE\\\label{Xi2a}&=1+\!\!\sum_{N=1}^\infty z^N\!\!\int_0^\infty\sum_{k=1}^N f(k,E) g(E,N)e^{-\beta E}dE.
\end{align}
To proceed further we have to assume that \cite{bookRudin}: i. the sum in Eq.~(\ref{af1}) is identically equal to one almost everywhere, that is, except on a set of measure zero, and ii. the unknown function, $f(k,E)$, multiplied by probability, $g(E,N)e^{-\beta E}$, that energy of
the considered system of $N$ particles lies in a narrow range between $E$ and
$E+dE$, is an essentially bounded function of $E$. In line with the two assumptions, the grand partition function, $\Xi_1$, can be rewritten as
\begin{equation}\label{Xi3}
\Xi_1(\beta,z)=1+\sum_{N=1}^\infty z^N\sum_{k=1}^N\int_0^\infty\!\!f(k,E)g(E,N)e^{-\beta E}dE.
\end{equation}

The grand partition function, $\Xi_2(\beta,z)$, may also be written in the
following way
\begin{equation}\label{Xi4}
\Xi_2(\beta,z)=e^{-\beta\Phi(\beta, z)}= \exp\left[-\beta\sum_{m=0}^\infty\frac{z^m}{m!}\phi_m(\beta)\right],
\end{equation}
where $\Phi(\beta,z)$ is the grand thermodynamic potential, derivatives of
which, $\phi_m(\beta)$, are denoted by
\begin{equation}\label{Xi4a}
\phi_m(\beta)=\left.\frac{\partial^m\Phi}{\partial z^m}\right|_{z=0},
\end{equation}
with
\begin{equation}\label{Xi4b}
\phi_0(\beta)=0,
\end{equation}
which results from simple comparison of Eqs.~(\ref{Xi1}) and~(\ref{Xi4}) for
$z=0$. Now, taking advantage of the generating function for partial Bell polynomials, $B_{N,k}(\{\phi_n\})$, which are the polynomials in an infinite number of variables
$\{\phi_n\}=\phi_1,\phi_2,\phi_3\dots$, defined by the formal double series expansion \cite{bookComtet}
\begin{eqnarray}\label{genbell1}
G(t,u)&=& \exp\left[u\sum_{m=1}^\infty\phi_m\frac{t^m}{m!}\right]\\\label{genbell2}&=& 1+\sum_{N=1}^\infty\frac{t^N}{N!}\sum_{k=1}^Nu^kB_{N,k}(\{\phi_n\}),
\end{eqnarray}
Eq.~(\ref{Xi4}) can be rewritten as
\begin{equation}\label{Xi5}
\Xi_2(\beta,z)=1+\sum_{N=1}^\infty \frac{z^N}{N!}\sum_{k=1}^N(-\beta)^kB_{N,k}(\{\phi_n(\beta)\}),
\end{equation}
where $B_{N,k}(\{\phi_n\})=B_{N,k}(\phi_1,\phi_2,\dots,\phi_{N-k+1})$
represent the so-called partial (or incomplete) Bell polynomials, which are
defined as
\begin{equation}\label{Bell}
B_{N,k}(\{\phi_n\})=N!\sum\prod_{n=1}^{N-k+1}\frac{1}{c_n!}\left(\frac{\phi_n}{n!}\right)^{c_n},
\end{equation}
where the summation takes place over all integers $c_n\geq 0$, such that
$\sum_{n}c_n=k$ and $\sum_{n}nc_n=N$.

Finally, comparing Eqs.~(\ref{Xi3}) and (\ref{Xi5}) one can readily obtain
the following expression
\begin{equation}\label{DOS00}
\sum_{k=1}^N\int_0^{\infty}f(k,E)g(E,N)e^{-\beta E}dE=\sum_{k=1}^N\frac{(-\beta)^k}{N!}B_{N,k}(\{\phi_n\}).
\end{equation}
Given combinatorial meaning of Bell polynomials (see Sec.~\ref{SecBell}) and also having in mind physical meaning of the probability $f(k,E)$, one can find that the equality of the two series in Eq.~(\ref{DOS00}) implies equality of their terms (see Sec.~\ref{SecIT}), i.e.
\begin{equation}\label{DOS0}
\int_0^{\infty}f(k,E)g(E,N)e^{-\beta E}dE=\frac{(-\beta)^k}{N!}B_{N,k}(\{\phi_n\}). \end{equation}
Using properties of Bell polynomials (see p.~135 in \cite{bookComtet}),
\begin{equation}\label{BellA}
a^kb^NB_{N,k}(\{\phi_n\})=B_{N,k}(\{ab^n\phi_n\}),
\end{equation}
Eq.~(\ref{DOS0}) can be simplified to
\begin{equation}\label{DOS1}
\int_0^{\infty}f(k,E)g(E,N)e^{-\beta E}dE=\frac{1}{N!}B_{N,k}(\{w_n\}),
\end{equation}
where
\begin{equation}\label{wn}
w_n=-\beta\phi_n.
\end{equation}

Eq.~(\ref{DOS1}) for the integral transform of $g(E,N)e^{-\beta E}dE$, with the kernel function given by $f(k,E)$, is the main result of this paper. The formula describes probability that a general system of $N$ particles at temperature $T=(k_B\beta)^{-1}$, regardless of its energy, consists of $k$ weakly coupled clusters. Furthermore, it provides a neat probabilistic description of microscopic arrangements (in terms of clusters) of the system investigated.

In the next section, a few points about Eq.~(\ref{DOS1}) are picked up. In particular, a novel and strictly microscopic interpretation of the grand thermodynamic potential is discussed and a new general combinatorial formula for the density of states is derived. The case of an ideal gas is considered as an example.

\section{Discussion of Eq.~(\ref{DOS1})}\label{sec2}

\subsection{Bell polynomials}\label{SecBell} The essential difficulty with Eq.~(\ref{DOS1}) arises from interpretation of Bell polynomials, $B_{N,k}(\{w_n\})$, which are described by Eq.~(\ref{Bell}). For this reason, we begin by elucidating their combinatorial meaning \cite{bookWilf, bookPitman}.

Suppose that $N$ particles labelled by elements of the set
$[N]=\{1,2,\dots,N\}$ are partitioned into clusters in such a way that each
particle belongs to a unique cluster. Obviously, for each composition of
$[N]$ into  $k$ disjoint non-empty clusters (subsets, blocks) of $n_i>0$
elements each, there are exactly
\begin{equation}\label{BellB}
{N\choose n_1,n_2,\dots n_k}=\frac{N!}{n_1!n_2!\dots n_k!}
\end{equation}
such compositions (partitions, distributions), where $n_1+n_2+\dots+n_k=N$.
Suppose further that each cluster of size $n$ in such a composition can be in any one of $w_n$ different internal states. Then, the number of configurations of the system of $N$ particles with $k$ clusters is given by
\begin{equation}\label{BellC}
N!\prod_{n=1}^{N-k+1}\left(\frac{w_n}{n!}\right)^{c_n},
\end{equation}
where $c_n\geq 0$ stands for the number of clusters of size $n$,
the largest cluster size is $N-k+1$, and
\begin{equation}\label{BellD}
\sum_{n=1}^{N-k+1}c_n=k,\;\;\;\;\;\;\;\; \sum_{n=1}^{N-k+1}nc_n=N.
\end{equation}

Eq.~(\ref{BellC}) describes system of $N$ distinguishable particles with $k$
clusters. If one assumes that the considered particles are indistinguishable,
and also that clusters of the same size are indistinguishable, the number of
configurations (compositions) becomes
\begin{equation}\label{BellE}
\prod_{n=1}^{N-k+1}\frac{1}{c_n!}\left(\frac{w_n}{n!}\right)^{c_n},
\end{equation}
which corresponds to Eq.~(\ref{BellC}) divided by $N!\prod_nc_n!$.

Finally, summing Eq.~(\ref{BellE}) over all integers $c_n\geq 0$ specified by
Eqs.~(\ref{BellD}) one gets the right-hand side of our main result, Eq.~(\ref{DOS1}).

\subsection{How Eq.~(\ref{DOS00}) implies Eq.~(\ref{DOS1})}\label{SecIT}

A pivotal role in our reasoning is played by the probability distribution $f(k,E)$ that the considered system of $N$ particles having energy $E$ consists of $k$ non-interacting, disjoint, and statistically independent clusters. In the simplest case of $N=1$, Eq.~(\ref{DOS00}) is equivalent to Eq.~(\ref{DOS1}), i.e.
\begin{equation}\label{N1}
\int_0^{\infty}f(1,E)g(E,1)e^{-\beta E}dE= B_{1,1}(\{w_n\})=w_1.
\end{equation}
In words, the last expression describes probability of the system which consists of the only one particle. Obviously, in this case the particle itself is the only cluster of size $n=1$, and its number of internal states is $w_1$.

In the case of $N=2$, Eq.~(\ref{DOS00}) can be written in the following way
\begin{eqnarray}\nonumber
\int_0^{\infty}\!\!f(1,E)g(E,2)e^{-\beta E}\!dE\!&\!+\!&\!\int_0^{\infty}\!\!f(2,E)g(E, 2)e^{-\beta E}\!dE\\
\label{N2b}
=\frac{1}{2}B_{2,1}(\{w_n\})&+&\frac{1}{2}B_{2,2}(\{w_n\})\\
\nonumber=w_2&+&w_1^2.
\end{eqnarray}
In accordance with Eq.~(\ref{N1}), the second terms in each row of the above three-line expression
represent thermodynamic probability of two, $k=2$, disjoint and statistically independent clusters of size $n=1$, i.e.
\begin{eqnarray}\nonumber w_1^2\!&\!=\!&\!\frac{1}{2}B_{2,2}(\{w_n\})=\left(\int_0^{\infty}f(1,E)g(E,1)e^{-\beta E}dE\right)^2\\\label{N2k2}&=&\int_0^{\infty}f(2,E)g(E, 2)e^{-\beta E}dE,
\end{eqnarray}
where we have exploited a useful feature of the grand canonical ensemble that, for non-interacting particles, the grand partition function factorizes into a product of grand partition functions for each single particle.

Therefore, since two particles, $N=2$, may be placed in either two clusters of size $n=1$, or a single one of size $n=2$, the first terms in Eqs.~(\ref{N2b}) should stand for thermodynamic probability of the second realizability, i.e. they must describe the number of microstates of the system with the only cluster of size $n=2$,
\begin{eqnarray}\label{N2k1}
\int_0^{\infty}f(1,E)g(E, 2)e^{-\beta E}dE=\frac{1}{2}B_{2,1}(\{w_n\})=w_2.
\end{eqnarray}
Similarly to what has been said about the the right-hand-side of Eq.~(\ref{N1}), it is clear the right-hand-side of the last equation, $w_2$, is the number of internal states of a single cluster of size $n=2$.

Analogously, one can show that Eq.~(\ref{DOS1}) is true for all $k\geq 3$.

\subsection{Microscopic interpretation of the grand thermodynamic potential}

From the previous subsection it is clear that the right-hand side of Eq.~(\ref{DOS1}) represents the number of microscopic arrangements of a system of $N$ particles at temperature $T$ which consists of $k$ clusters of various sizes. The number of internal states, $w_n$, of the cluster of size $n$ is, up to a multiplicative constant, $-\beta$, proportional to the $n$th derivative of the grand potential, $\phi_n(\beta)$, cf.~Eq.~(\ref{wn}).

The understanding of the grand potential as the exponential generating function,
\begin{equation}\label{LandauFE}
\Phi(z,\beta)=\sum_{n=1}^\infty\frac{z^n}{n!}\phi_n(\beta),
\end{equation}
for the number of internal states of $n$-clusters (i.e. groups of $n$ interacting particles) is novel. It offers a completely new microscopic interpretation of this well-known thermodynamic function.

It is also interesting to note that from the microscopic interpretation of the grand potential, very general implications for the stability requirements on the Landau free energy arise. In general, the classical stability theory (see Chap.~2H in \cite{bookReichl}, and Chaps.~12-14 in \cite{bookKondepudi}) places certain conditions on the sign of second derivatives of thermodynamic potentials of the system investigated. However, given our derivations it seems necessary to expand the Gibbs stability requirements to higher derivatives. To be definite, from the non-negative definiteness of the coefficients, $w_n$, of Bell polynomials in Eq.~(\ref{DOS1}), one gets the following conditions 
\begin{equation}\label{stability}
\forall_{n\geq1}\;\;\phi_n(\beta)\leq 0.
\end{equation}
One can speculate that the requirements may have connections with higher order phase transitions (in a similar way as the classical stability theory, which exploits second order derivatives, is used to characterize critical phenomena). The analogy will be further elaborated in our subsequent papers.

\subsection{The case of the classical ideal gas}\label{SecIdeal}

In this section, we apply our formalism to the classical ideal gas. In this case, the grand partition function can be written as, cf.~Eq.~(\ref{Xi1}),
\begin{equation}\label{ideal1}
\Xi_G(\beta,z)=1+\sum_{N=1}^\infty\frac{z^N}{N!}(Z_1(\beta))^N=e^{zZ_1(\beta)},
\end{equation}
where
\begin{equation}\label{ideal2}
Z_1(\beta)=\frac{V}{\lambda^3}=V\left(\frac{2\pi m}{h^2\beta}\right)^{3/2}
\end{equation}
is the partition function of a single particle, $\lambda$ stands for the de~Broglie wavelength, $V$ is volume, and $h$ is the Planck's constant.

Using Eq.~(\ref{Xi4}) one finds that the grand potential of the gas is a linear function of the parameter $z$,
\begin{equation}\label{ideal3}
\Phi_G(\beta,z)=-\frac{Z_1(\beta)}{\beta}z,
\end{equation}
and the coefficients of the Bell polynomial in Eq.~(\ref{DOS1}) are equal
\begin{equation}\label{ideal4}
w_1=Z_1(\beta)\;\;\;\;\;\mbox{and}\;\;\;\;\;\forall_{n\geq2}\;\;w_n=0.
\end{equation}

From Eqs.~(\ref{ideal4}) it follows that the only possible cluster decomposition of a general system of $N$ particles described by the linear grand potential is the one in which each particle is located in a separate cluster. In fact, in the considered case of the ideal gas, where particles do not interact with each other, the result is the only reasonable. Furthermore, since the number of internal states of a single particle is $Z_1=V\lambda^{-3}$, and the particles are statistically independent, one gets that thermodynamic probability of the system as a whole is $Z_1^N=w_1^N=B_{N,N}(\{w_n\})$, which agrees with Eq.~(\ref{DOS1}).

\subsection{The integral transform for the density of states}

Although in Eq.~(\ref{DOS1}), the kernel function, $f(k,E)$, which describes the probability of the considered system of $N$ particles to consist of $k$ weakly coupled clusters, depends on the system itself \footnote{In the case of the ideal gas, one has $f(k,E)=\delta_{kN}$, where $\delta_{ij}$ is the Kronecker delta.}, a given explicitly $f(k,E)$ may also provide valuable information about the system.

For example, let the kernel function be the Poisson distribution in the number of clusters,
\begin{equation}\label{poisson0}
f(k,E)\simeq P(k;\langle k\rangle)=\frac{e^{-\langle k\rangle}\langle k\rangle^k}{k!},
\end{equation}
where $k=E/q$, $\langle k\rangle=\langle E\rangle/q$, and $q$ stands for an energy quanta, such that \footnote{Eq.~(\ref{poisson1}) depends on $\langle k\rangle$. However, in most of interesting systems, due to small energy fluctuations in equilibrium ensembles, there must exist such a value of $q$ for which the equation is well-fulfilled in the energy range, where the corresponding probability of macrostates is meaningful.}
\begin{equation}\label{poisson1}
\sum_{k=1}^{N}P(k;\langle k\rangle)\simeq \sum_{k=0}^{\infty}P(k;\langle k\rangle)=1.
\end{equation}
Inserting Eq.~(\ref{poisson0}) into Eq.~(\ref{DOS1}) one gets the following integral formula for the density of states
\begin{equation}\label{poissonDOS0}
P.T.\left[g(E,N)e^{-\beta E};\frac{E}{q},k\right]=\frac{1}{N!}B_{N,k}(\{-\beta\phi_n\}),
\end{equation}
where the symbolic notation $P.T.[\dots]$ represents the Poisson transform defined as \cite{Wolf1964,Hindin1968}
\begin{equation}\label{PT}
P.T.[f(x);x,k]=\int_0^\infty\frac{e^{-x}x^k}{k!}f(x)dx=F(k).
\end{equation}

To elucidate how Eq.~(\ref{poissonDOS0}) can be used to describe energy distribution in the system composed of $N$ interacting particles, one should start with some remarks about the Poisson transform. Thus, when working with this transform it is important to understand, how it acts on arbitrary function. In some sense, it is reasonable to say that the transform, $F(k)$, is a kind of weighted moving average of the original function, $f(x)$. In comparison with the simple moving average, in which the unweighted mean over some range of neighboring values is taken into account, in the Poisson transform the averaging is performed over the whole domain, $x\in\langle 0,\infty)$, of the original function, $f(x)$, with the weights given by the Poisson distribution with the moving mean value which is equal to $x$. For this reason, due to properties of the Poisson distribution, the averaging is actually done over the nearest neighborhood of $x$. The transform, $F(k)=P.T.[f(x);x,k]$, looks like a smooth (or fuzzy) image of the original, $f(x)$, and it is often reasonable to assume that \footnote{In fact, the approximation is only acceptable when the function varies slowly enough. The theorem justifying approximating function by its Poisson transform is given in Appendix B1 in Ref.\cite{Fronczak2010}. See also Refs.~\cite{Wolf1964,Fronczak2006}, in which exact formulas for the inverse Poisson transforms were derived.} $F(k)\simeq f(x)|_{x=k}$.

Therefore, simplifying, one could say that the left-hand-side of Eq.~(\ref{poissonDOS0}) represents a smoothed thermodynamic probability,
\begin{equation}\label{poissonDOS2}
g(E,N)e^{-\beta E}|_{E=kq}\simeq \frac{1}{N!}B_{N,k}(\{w_n\}),
\end{equation}
from which one gets the approximated general combinatorial formula for the density of states
\begin{equation}\label{poissonDOS3}
g(E,N)|_{E=kq}\simeq \frac{e^{\beta qk}}{N!}B_{N,k}(\{-\beta\phi_n\}),
\end{equation}
where $q$ may be interpreted as the resolution parameter for the energy distribution.

\section{Summary and concluding remarks}

In this paper we have derived an exact combinatorial formula for probability that a general system of $N$ particles at temperature $T$ consists of $k$ weakly-coupled clusters of various sizes. We have also found that the grand potential (i.e. the Landau free energy) may be considered as the exponential generating function for the number of internal states of these clusters. Finally, we have derived an approximated expression for the density of states.

We believe that the approach, when applied to systems, such as lattice gases  will provide important insights into mechanisms that lead to the occurrence of phase transitions. We also believe that the approach is the correct starting point for a future theory of higher order phase transitions.

\acknowledgments
I would like to thank my husband, Dr. Piotr Fronczak, for his wide-ranging support and patience. I also wish to thank MSc. Grzegorz Siudem, a brilliant PhD student graduated in both physics and mathematics, for clarifying some vagueness in definition of the kernel function, $f(k,E)$. Finally I thank Prof. Zdzis\l aw Burda for the valuable discussion on fundamentals of statistical physics during Summer Solstice Conference on Discrete Models of Complex Systems in Gda\'nsk (Poland) in June
2009. The work has been supported from the internal funds of the Faculty of
Physics at Warsaw University of Technology and from the Ministry of Science
and Higher Education in Poland (national three-year scholarship for
outstanding young scientists 2010).


\begin{thebibliography}{25}
\expandafter\ifx\csname natexlab\endcsname\relax\def\natexlab#1{#1}\fi
\expandafter\ifx\csname bibnamefont\endcsname\relax
  \def\bibnamefont#1{#1}\fi
\expandafter\ifx\csname bibfnamefont\endcsname\relax
  \def\bibfnamefont#1{#1}\fi
\expandafter\ifx\csname citenamefont\endcsname\relax
  \def\citenamefont#1{#1}\fi
\expandafter\ifx\csname url\endcsname\relax
  \def\url#1{\texttt{#1}}\fi
\expandafter\ifx\csname urlprefix\endcsname\relax\def\urlprefix{URL }\fi
\providecommand{\bibinfo}[2]{#2}
\providecommand{\eprint}[2][]{\url{#2}}

\bibitem[{\citenamefont{Ursell}(1927)}]{1927Ursell}
\bibinfo{author}{\bibfnamefont{H.~D.} \bibnamefont{Ursell}},
  \bibinfo{journal}{Proc. Cambridge Philos. Soc.}
  \textbf{\bibinfo{volume}{23}}, \bibinfo{pages}{685} (\bibinfo{year}{1927}).

\bibitem[{\citenamefont{Mayer}(1937)}]{1937MayerJChemPhys1}
\bibinfo{author}{\bibfnamefont{J.~E.} \bibnamefont{Mayer}},
  \bibinfo{journal}{J. Chem. Phys.} \textbf{\bibinfo{volume}{5}},
  \bibinfo{pages}{67} (\bibinfo{year}{1937}).

\bibitem[{\citenamefont{Mayer and Ackermann}(1937)}]{1937MayerJChemPhys2}
\bibinfo{author}{\bibfnamefont{J.~E.} \bibnamefont{Mayer}} \bibnamefont{and}
  \bibinfo{author}{\bibfnamefont{P.~G.} \bibnamefont{Ackermann}},
  \bibinfo{journal}{J. Chem. Phys.} \textbf{\bibinfo{volume}{5}},
  \bibinfo{pages}{74} (\bibinfo{year}{1937}).

\bibitem[{\citenamefont{Mayer and Harrison}(1938)}]{1938MayerJChemPhys}
\bibinfo{author}{\bibfnamefont{J.~E.} \bibnamefont{Mayer}} \bibnamefont{and}
  \bibinfo{author}{\bibfnamefont{S.~F.} \bibnamefont{Harrison}},
  \bibinfo{journal}{J. Chem. Phys.} \textbf{\bibinfo{volume}{6}},
  \bibinfo{pages}{87} (\bibinfo{year}{1938}).

\bibitem[{\citenamefont{Mayer and Montroll}(1941)}]{1941MayerJChemPhys}
\bibinfo{author}{\bibfnamefont{J.~E.} \bibnamefont{Mayer}} \bibnamefont{and}
  \bibinfo{author}{\bibfnamefont{E.}~\bibnamefont{Montroll}},
  \bibinfo{journal}{J. Chem. Phys.} \textbf{\bibinfo{volume}{9}},
  \bibinfo{pages}{2} (\bibinfo{year}{1941}).

\bibitem[{\citenamefont{Mayer and Goeppert-Mayer}(1977)}]{bookMayer}
\bibinfo{author}{\bibfnamefont{J.~E.} \bibnamefont{Mayer}} \bibnamefont{and}
  \bibinfo{author}{\bibfnamefont{M.}~\bibnamefont{Goeppert-Mayer}}
  (\bibinfo{publisher}{Wiley}, \bibinfo{address}{New York},
  \bibinfo{year}{1977}), chap.~\bibinfo{chapter}{8}, pp.
  \bibinfo{pages}{222--289}, \bibinfo{edition}{2nd} ed.

\bibitem[{\citenamefont{Hill}(1956)}]{bookHill}
\bibinfo{author}{\bibfnamefont{T.~L.} \bibnamefont{Hill}}
  (\bibinfo{publisher}{McGraw-Hill}, \bibinfo{address}{New York},
  \bibinfo{year}{1956}), chap.~\bibinfo{chapter}{5}, pp.
  \bibinfo{pages}{122--178}, \bibinfo{edition}{2nd} ed.

\bibitem[{\citenamefont{Reichl}(1998)}]{bookReichl}
\bibinfo{author}{\bibfnamefont{L.~E.} \bibnamefont{Reichl}},
  \emph{\bibinfo{title}{A Modern Course in Statistical Physics}}
  (\bibinfo{publisher}{Wiley}, \bibinfo{address}{New York},
  \bibinfo{year}{1998}).

\bibitem[{\citenamefont{Stanley}(1997)}]{bookStanley}
\bibinfo{author}{\bibfnamefont{R.~P.} \bibnamefont{Stanley}},
  \emph{\bibinfo{title}{Enumerative Combinatorics, vol. 1.}}
  (\bibinfo{publisher}{Cambridge University Press},
  \bibinfo{address}{Cambridge}, \bibinfo{year}{1997}).

\bibitem[{\citenamefont{Comtet}(1974)}]{bookComtet}
\bibinfo{author}{\bibfnamefont{L.}~\bibnamefont{Comtet}}
  (\bibinfo{publisher}{Reidel Publishing Company},
  \bibinfo{address}{Dordrecht}, \bibinfo{year}{1974}), pp.
  \bibinfo{pages}{133--137}.

\bibitem[{\citenamefont{Wilf}(1990)}]{bookWilf}
\bibinfo{author}{\bibfnamefont{H.~S.} \bibnamefont{Wilf}},
  \emph{\bibinfo{title}{Generatingfunctionology}} (\bibinfo{publisher}{Academic
  Press}, \bibinfo{address}{New York}, \bibinfo{year}{1990}).

\bibitem[{\citenamefont{Born and Fuchs}(1938)}]{1938Born}
\bibinfo{author}{\bibfnamefont{M.}~\bibnamefont{Born}} \bibnamefont{and}
  \bibinfo{author}{\bibfnamefont{K.}~\bibnamefont{Fuchs}},
  \bibinfo{journal}{Proc. Roy. Soc. (London)} \textbf{\bibinfo{volume}{A166}},
  \bibinfo{pages}{391} (\bibinfo{year}{1938}).

\bibitem[{\citenamefont{Kahn and Uhlenbeck}(1938)}]{1938Kahn}
\bibinfo{author}{\bibfnamefont{B.}~\bibnamefont{Kahn}} \bibnamefont{and}
  \bibinfo{author}{\bibfnamefont{G.~E.} \bibnamefont{Uhlenbeck}},
  \bibinfo{journal}{Physica} \textbf{\bibinfo{volume}{5}}, \bibinfo{pages}{399}
  (\bibinfo{year}{1938}).

\bibitem[{\citenamefont{Riddel and Uhlenbeck}(1953)}]{1953Riddel}
\bibinfo{author}{\bibfnamefont{R.~J.} \bibnamefont{Riddel}} \bibnamefont{and}
  \bibinfo{author}{\bibfnamefont{G.~E.} \bibnamefont{Uhlenbeck}},
  \bibinfo{journal}{J. Chem. Phys.} \textbf{\bibinfo{volume}{21}},
  \bibinfo{pages}{2056} (\bibinfo{year}{1953}).

\bibitem[{\citenamefont{Yang and Lee}(1952)}]{1952YangPhysRev}
\bibinfo{author}{\bibfnamefont{C.~N.} \bibnamefont{Yang}} \bibnamefont{and}
  \bibinfo{author}{\bibfnamefont{T.~D.} \bibnamefont{Lee}},
  \bibinfo{journal}{Phys. Rev.} \textbf{\bibinfo{volume}{87}},
  \bibinfo{pages}{404} (\bibinfo{year}{1952}).

\bibitem[{\citenamefont{Whittle}(1986)}]{bookWhittle}
\bibinfo{author}{\bibfnamefont{P.}~\bibnamefont{Whittle}},
  \emph{\bibinfo{title}{Systems in Stochastic Equilibrium}}
  (\bibinfo{publisher}{Wiley}, \bibinfo{year}{1986}).

\bibitem[{\citenamefont{Pitman}(2006)}]{bookPitman}
\bibinfo{author}{\bibfnamefont{J.}~\bibnamefont{Pitman}},
  \emph{\bibinfo{title}{Combinatorial Stochastic Processes}}
  (\bibinfo{publisher}{Springer}, \bibinfo{year}{2006}), \bibinfo{note}{lecture
  notes in mathematics Ecole d'Et\'{e} de probabilit\'{e}s de Saint-Flour
  XXXII-2002. Vol. 1875}.

\bibitem[{\citenamefont{Berestycki and Pitman}(2007)}]{2007BerestyckiJStatPhys}
\bibinfo{author}{\bibfnamefont{N.}~\bibnamefont{Berestycki}} \bibnamefont{and}
  \bibinfo{author}{\bibfnamefont{J.}~\bibnamefont{Pitman}},
  \bibinfo{journal}{J. Stat. Phys.} \textbf{\bibinfo{volume}{127}},
  \bibinfo{pages}{381} (\bibinfo{year}{2007}).

\bibitem[{\citenamefont{Pathria and Beale}(2011)}]{bookPathria}
\bibinfo{author}{\bibfnamefont{R.~K.} \bibnamefont{Pathria}} \bibnamefont{and}
  \bibinfo{author}{\bibfnamefont{P.~T.} \bibnamefont{Beale}}
  (\bibinfo{publisher}{Elsevier}, \bibinfo{address}{New York},
  \bibinfo{year}{2011}), chap.~\bibinfo{chapter}{10}, pp.
  \bibinfo{pages}{299--343}, \bibinfo{edition}{3rd} ed.

\bibitem[{\citenamefont{Rudin}(1991)}]{bookRudin}
\bibinfo{author}{\bibfnamefont{W.}~\bibnamefont{Rudin}},
  \emph{\bibinfo{title}{Functional Analysis}}
  (\bibinfo{publisher}{McGraw-Hill}, \bibinfo{year}{1991}).

\bibitem[{\citenamefont{Kondepudi and Prigogine}(1998)}]{bookKondepudi}
\bibinfo{author}{\bibfnamefont{D.}~\bibnamefont{Kondepudi}} \bibnamefont{and}
  \bibinfo{author}{\bibfnamefont{I.}~\bibnamefont{Prigogine}},
  \emph{\bibinfo{title}{Modern Thermodynamics: From Heat engines to Dissipative
  Structures}} (\bibinfo{publisher}{Wiley}, \bibinfo{year}{1998}).

\bibitem[{\citenamefont{Wolf and Mehta}(1964)}]{Wolf1964}
\bibinfo{author}{\bibfnamefont{E.}~\bibnamefont{Wolf}} \bibnamefont{and}
  \bibinfo{author}{\bibfnamefont{C.~L.} \bibnamefont{Mehta}},
  \bibinfo{journal}{Phys. Rev. Lett.} \textbf{\bibinfo{volume}{13}},
  \bibinfo{pages}{705} (\bibinfo{year}{1964}).

\bibitem[{\citenamefont{Hindin}(1968)}]{Hindin1968}
\bibinfo{author}{\bibfnamefont{H.~J.} \bibnamefont{Hindin}},
  \bibinfo{howpublished}{Conference record of the 2nd Asilomar Conference on
  Circuits and Systems, pp. 525--529, Electronics Engineers Inc., New York}
  (\bibinfo{year}{1968}).

\bibitem[{\citenamefont{Fronczak and Fronczak}(2010)}]{Fronczak2010}
\bibinfo{author}{\bibfnamefont{A.}~\bibnamefont{Fronczak}} \bibnamefont{and}
  \bibinfo{author}{\bibfnamefont{P.}~\bibnamefont{Fronczak}},
  \bibinfo{journal}{Phys. Rev. E} \textbf{\bibinfo{volume}{81}},
  \bibinfo{pages}{066112} (\bibinfo{year}{2010}).

\bibitem[{\citenamefont{Fronczak and Fronczak}(2006)}]{Fronczak2006}
\bibinfo{author}{\bibfnamefont{A.}~\bibnamefont{Fronczak}} \bibnamefont{and}
  \bibinfo{author}{\bibfnamefont{P.}~\bibnamefont{Fronczak}},
  \bibinfo{journal}{Phys. Rev. E} \textbf{\bibinfo{volume}{74}},
  \bibinfo{pages}{026121} (\bibinfo{year}{2006}).

\end{thebibliography}

\end{document}